\documentclass[preprint,prd,showpacs,showkeys,byrevtex,footinbib,eqsecnum,unsortedaddress]{revtex4}
\usepackage{amsfonts}
\usepackage{amsmath}
\usepackage{amssymb}

\input{tcilatex}

\begin{document}

\title{Ultrasoft Quark Damping in Hot QCD}
\author{A. Abada}
\email{a.abada@uaeu.ac.ae}
\affiliation{Physics Department, United Arab Emirates University, P.O.B. 17551, Al Ain,
United Arab Emirates}
\altaffiliation{On leave from: D\'{e}partement de Physique, ENS, BP 92 Vieux Kouba, 16050
Alger, Algeria.}
\author{N. Daira-Aifa}
\email{daira\_aifa@yahoo.fr}
\affiliation{Facult\'{e} de Physique, Universit\'{e} des Sciences et de la Technologie
Houari Boumediene, BP\ 32 El-Alia, Bab Ezzouar 16111 Alger,
Algeria\smallskip }
\author{K. Bouakaz}
\email{bouakazk@caramail.com}
\affiliation{D\'{e}partement de Physique, Ecole Normale Sup\'{e}rieure, BP 92 Vieux
Kouba, 16050 Alger, Algeria}

\begin{abstract}
We determine the quark damping rates in the context of next-to-leading order
hard-thermal-loop summed perturbation of high-temperature QCD where weak
coupling is assumed. The quarks are ultrasoft. Three types of divergent
behavior are encountered: infrared, light-cone and at specific points
determined by the gluon energies. The infrared divergence persists and is
logarithmic whereas the two others are circumvented.
\end{abstract}

\date{\today }
\maketitle

\section{Introduction}

RHIC results seem to indicate that in the temperature range between one to
two $T_{c}\simeq 200\mathrm{MeV}$, the critical temperature of deconfinement
from conventional hadronic states, there is an intermediary phase in which
the deconfined quarks and gluons remain strongly interacting \cite{shuryak1}%
, a scenario introduced in \cite{shuryak-zahed1,shuryak-zahed2,shuryak2}.
Since in this temperature regime the quark binding energy is about 4 GeV,
one order of magnitude higher than the temperature itself, perturbative
treatment based on quasi-free quarks and gluons is not adequate. Rather, a
hydrodynamic description of a near-perfect liquid seems to work and the
picture emerging is that of non-conventional bound states of quarks and
gluons: diquarks $qq$, baryons $qqq$, three-gluon bound states $ggg$ and
polymeric chains $\bar{q}g\dots gq$. A model of a strongly interacting
classical chromo-electric plasma can account of much of the phenomenological
results \cite{shuryak1}.

But the\ same quark binding energy decreases as the temperature increases
away from $T_{c}$ \cite{shuryak1}, and here, starting from about $3T_{c}$,
weak coupling between individual constituents may be expected. This is also
supported by the behavior of particle susceptibilities: baryons and other
bound states contribute in the intermediary temperature range, but only
quarks survive in the high-temperature limit \cite{liao-shuryak}.

Therefore, as the temperature increases, the picture emerging is that of a
hadronic phase with very short inter-quark binding lengths changing at $%
T_{c} $ into the so-called strongly coupled quark-gluon plasma (sQGP), a
near-perfect liquid phase of quark and gluon bound states with longer
inter-quasiparticle binding lengths. As the temperature increases further,
the sQGP changes into a weakly coupled quark-gluon plasma (wQGP) of quasi
quarks and gluons with chromo-neutralizing isotropic Debye clouds \cite%
{shuryak1}.

If perturbative QCD is to apply in the wQGP phase, the problem becomes: how
to organize it? It is known for some time that the standard loop-expansion
would break at some order, depending on the quantity of interest\ under
consideration \cite{linde1,linde2,gross-pisarski-yaffe}. It has also proved
inadequate when describing slow-moving particles since it does not reflect
an expansion in powers of the coupling \cite%
{kalash-klimo,klimov1,klimov2,weldon1,weldon2}. Regarding this last
difficulty, an important improvement has been the dressing of the
lowest-order propagators and vertices with the so-called hard thermal loops
(HTL) \cite{BP1,BP2,BP3,pisarski1,pisarski2,pisarski3,frenkel-taylor,le
bellac}. Since then, work has flourished and many phenomenological aspects
of the presumed wQGP as well as more theoretical aspects of high-temperature
QCD have been addressed in the context of lowest and next-to-leading orders
of the so-called HTL-perturbation theory \cite{kraemmer-rebhan}. But
HTL-dressed perturbative QCD is not itself safe from chronic problems, one
important one being the non-screening of static chromo-magnetic fields at
lowest order which plagues the theory with infrared divergences in
next-to-leading order calculations. The determination of the chromo-magnetic
correlation length may not even be accessible perturbatively \cite%
{kraemmer-rebhan}.

The present work finishes a calculation started in \cite{ABD}. The context
is next-to-leading order HTL-dressed perturbative QCD and weak coupling is
assumed. The aim is to complete the determination of the quark damping rates
to second order in the ultrasoft external momentum. The coefficients of
zeroth order are the ones already found in \cite%
{kobes-kunstatter-mak,braaten-pisarski (quarks)}; they are finite and
positive. The coefficients of first order we find are also finite and free
from any divergence, but the coefficients of second order are
logarithmically divergent in the infrared. This work is meant to be an
additional contribution investigating the analytic infrared behavior of QCD
at high-temperature. It is certainly more pressing nowadays to have a better
theoretical understanding of the sQGP phase of hadronic matter, but
ultimately, two issues have to be addressed: (i) Must we look for a quantum
field theory description of the different phases of quarks and gluons? (ii)
Do we want to reproduce the eventual phase transitions in this context?

This article is organized as follows. After this introduction, section two
recalls the essential results of \cite{ABD} to which the present work is a
follow-up. The analytic expression of the damping rates is given in the form
of integrals over functions involving products of spectral distributions and
their first and second derivatives. Section three takes up from these
expressions and perform the integrals. The steps of the calculations are
detailed using a generic form. The occurrence of divergences is discussed.
Three types occur: infrared, light-cone, and divergences at specific points
determined by the gluon energies. The infrared divergences are extracted and
will stay, whereas the two other kinds are dealt away with. Section four
summarizes the work and finishes the article with concluding remarks.

\section{Quark damping rates in HTL-summed perturbation}

We consider a theory with $N_{c}$ colors and $N_{f}$ flavors. Imaginary-time
formalism is used throughout. The euclidean momentum of the quark is $P^{\mu
}=(p_{0},\mathbf{p})$ such that $P^{2}=p_{0}^{2}+p^{2}$ with $%
p_{0}=(2n+1)\pi T$, a fermionic Matsubara frequency; $n$ is an integer. Once
all intermediary steps are carried out, real-time amplitudes are obtained
via the analytic continuation $p_{0}=-i\omega +0^{+}$ where $\omega $ is the
energy of the quark. The temperature $T$ determines the hard scale, $gT$ the
soft scale where $g$ is the (weak) coupling constant, and $g^{2}T$ the
ultrasoft scale. An infrared cutoff $\eta \sim g^{2}T$ is introduced.

\subsection{Dressing the propagators and vertices}

The HTL-dressed quark propagator can be written as:%
\begin{equation}
^{\ast }\Delta _{F}\left( P\right) =-\left[ \mathbf{\gamma }_{+p}\,\Delta
_{+}\left( P\right) +\mathbf{\gamma }_{-p}\,\Delta _{-}\left( P\right) %
\right] .  \label{delta-f}
\end{equation}%
$\gamma ^{\mu }$ are the euclidean Dirac matrices, $\mathbf{\gamma }_{\pm
p}=\left( \gamma ^{0}\pm i\mathbf{\gamma }.\mathbf{\hat{p}}\right) /2$ and $%
\Delta _{\pm }=\left( D_{0}\mp D_{s}\right) ^{-1}$ with:%
\begin{equation}
D_{0}\left( P\right) =ip_{0}-\frac{m_{f}^{2}}{p}Q_{0}\left( \frac{ip_{0}}{p}%
\right) ;\qquad D_{s}\left( P\right) =p+\frac{m_{f}^{2}}{p}\left[ 1-\frac{%
ip_{0}}{p}Q_{0}\left( \frac{ip_{0}}{p}\right) \right] .  \label{D0-Ds}
\end{equation}%
$m_{f}=\sqrt{C_{f}/8}\,gT$ is the lowest-order quark thermal mass with $%
C_{f}=(N_{c}^{2}-1)/2N_{c}$, and $Q_{0}\left( x\right) =\frac{1}{2}\ln 
\dfrac{x+1}{x-1}$. The poles of $\Delta _{\pm }(-i\omega ,\mathbf{p})$
determine the dispersion laws $\omega _{\pm }(p)$ to lowest order in $g$.
The $+$ sign is for real quarks and the $-$ sign for the so-called
`plasminos' \cite{braaten}, thermally excited quasiparticles. For soft
quarks and using the notation $\bar{p}=p/m_{f}$, one has:%
\begin{equation}
\omega _{\pm }(p)=m_{f}\left[ 1\pm \frac{1}{3}\bar{p}+\frac{1}{3}\bar{p}%
^{2}\mp \frac{16}{135}\bar{p}^{3}+\frac{1}{54}\bar{p}^{4}\pm \frac{32}{2835}%
\bar{p}^{5}-\frac{139}{12150}\bar{p}^{6}\pm \mathcal{O}\left( \bar{p}%
^{7}\right) \right] .  \label{omega-pm}
\end{equation}

The quark damping rates $\gamma _{\pm }\left( p\right) $ are obtained by
including in the dispersion relations the HTL-dressed one-loop-order quark
self-energy $^{\ast }\Sigma (P)$. The inverse quark propagator becomes:%
\begin{equation}
\Delta _{F}^{-1}\left( P\right) =\,^{\ast }\Delta _{F}^{-1}\left( P\right)
-\,^{\ast }\Sigma (P)\,.  \label{full-propa-a}
\end{equation}%
The decomposition $^{\ast }\Sigma =\gamma ^{0}\,^{\ast }D_{0}+i\mathbf{%
\gamma }.\mathbf{\hat{p}\;}^{\ast }D_{s}$ implies:%
\begin{equation}
\Delta _{F}^{-1}\left( P\right) =-\left[ \gamma ^{0}\left( D_{0}+\,^{\ast
}D_{0}\right) +i\mathbf{\gamma }.\mathbf{\hat{p}}\left( D_{s}+\,^{\ast
}D_{s}\right) \right] .  \label{full-propa-b}
\end{equation}%
With the definitions $\gamma _{\pm }(p)\equiv -\func{Im}\Omega _{\pm }\left(
p\right) $ with $\Omega _{\pm }$ the poles of $\Delta _{F}(-i\Omega ,\mathbf{%
p})$, and since $^{\ast }\Sigma $ is $g$-times smaller than $^{\ast }\Delta
_{F}^{-1},$ we have to order $g^{2}T$:%
\begin{equation}
\gamma _{\pm }\left( p\right) =\left. \frac{\func{Im}\mathrm{\,}^{\ast
}f_{\pm }\left( -i\omega ,p\right) }{\partial _{\omega }f_{\pm }\left(
-i\omega ,p\right) }\right\vert _{\omega =\omega _{\pm }(p)+i0^{+}},
\label{gamma-pm-a}
\end{equation}%
where $f_{\pm }=D_{0}\mp D_{s}$, $^{\ast }f_{\pm }=\,^{\ast }D_{0}\mp
\,^{\ast }D_{s}$ and $\partial _{\omega }$ stands for $\partial /\partial
\omega $. Expanding the denominator in the above relation in powers of $\bar{%
p}$ using the expressions in (\ref{D0-Ds}), one obtains:%
\begin{equation}
\gamma _{\pm }\left( p\right) =\left. \frac{1}{2}\left[ 1\pm \frac{2}{3}\bar{%
p}-\frac{2}{9}\bar{p}^{2}\pm \mathcal{O}\left( \bar{p}^{3}\right) \right] \,%
\func{Im}\,^{\ast }f_{\pm }\left( -i\omega ,p\right) \right\vert _{\omega
=\omega _{\pm }\left( p\right) +i0^{+}}\,.  \label{gamma-pm-b}
\end{equation}%
Therefore, determining $\gamma _{\pm }\left( p\right) $ amounts to
calculating $\func{Im}\,^{\ast }\Sigma \left( P\right) $.

In HTL-summed perturbation \cite{BP1,frenkel-taylor,le bellac}, the
one-loop-order quark self-enegy writes:%
\begin{equation}
^{\ast }\Sigma \left( P\right) =\,^{\ast }\Sigma _{1}\left( P\right)
+\,^{\ast }\Sigma _{2}\left( P\right) .  \label{eff-self-energy}
\end{equation}%
The first contribution is:%
\begin{equation}
^{\ast }\Sigma _{1}\left( P\right) =-g^{2}C_{f}\mathrm{Tr}_{\mathrm{soft}}%
\left[ ^{\ast }\Gamma ^{\mu }\left( P,-Q;-K\right) \,^{\ast }\Delta
_{F}\left( Q\right) \,^{\ast }\Gamma ^{\nu }\left( -P,Q;K\right) \,^{\ast
}\Delta _{\mu \nu }\left( K\right) \right] ,  \label{eff-self-energy1}
\end{equation}%
a QED-like loop formed by two quark-gluon vertices connected by one gluon
propagator and one quark propagator. The second contribution is:%
\begin{equation}
^{\ast }\Sigma _{2}\left( P\right) =-\frac{i}{2}g^{2}C_{f}\mathrm{Tr}_{%
\mathrm{soft}}\left[ ^{\ast }\widetilde{\Gamma }^{\mu \nu }\left(
P,-P;K,-K\right) \,^{\ast }\Delta _{\mu \nu }\left( K\right) \right] ,
\label{eff-self-energy2}
\end{equation}%
a purely hard-thermal-loop two-gluon-quark-antiquark vertex with one gluon
propagator. In the above two expressions, $K$ is the soft loop momentum, $%
Q=P-K$ and $\mathrm{Tr}\equiv T\underset{k_{0}}{\sum }\int \frac{d^{3}k}{%
(2\pi )^{3}}$ with $k_{0}=2n\pi T$ when bosonic or $k_{0}=\left( 2n+1\right)
\pi T$ when fermionic. The subscript `soft' means that only soft values of $%
K $ are allowed in the integrals; hard values have already dressed the
propagators and vertices.

To be complete, we give the expressions of the gluon propagator and the
vertices involved in (\ref{eff-self-energy1}) and (\ref{eff-self-energy2}).
In the strict Coulomb gauge, $^{\ast }\Delta _{00}\left( K\right) =\,^{\ast
}\Delta _{l}\left( K\right) $, $^{\ast }\Delta _{0i}\left( K\right) =0$ and $%
^{\ast }\Delta _{ij}\left( K\right) =\left( \delta _{ij}-\hat{k}_{i}\hat{k}%
_{j}\right) \,^{\ast }\Delta _{t}\left( K\right) $ with $^{\ast }\Delta _{l}$
and $^{\ast }\Delta _{t}$ given by:%
\begin{equation}
^{\ast }\Delta _{l}\left( K\right) =\frac{1}{k^{2}-\delta \Pi _{l}\left(
K\right) }\,;\,\,\,\,\,\,\,\,\,\,\,\,\,\,\,\,\,\,\,\,^{\ast }\Delta
_{t}\left( K\right) =\frac{1}{K^{2}-\delta \Pi _{t}\left( K\right) }\,,
\label{propa-glu}
\end{equation}%
where $\delta \Pi _{l}(K)=3m_{g}^{2}Q_{1}(\frac{ik_{0}}{k})$ and $\delta \Pi
_{t}(K)=\frac{3}{5}m_{g}^{2}\left[ Q_{3}(\frac{ik_{0}}{k})-Q_{1}(\frac{ik_{0}%
}{k})-\frac{5}{3}\right] $ are the gluonic hard thermal loops. Here $Q_{i}(%
\frac{ik_{0}}{k})$ is a Legendre function of the second kind and $m_{g}=%
\sqrt{N_{c}+N_{f}/2}gT/3$ is the gluon thermal mass. Finally, the
HTL-dressed vertices $^{\ast }\Gamma $ are as follows:%
\begin{eqnarray}
^{\ast }\Gamma ^{\mu }(P,Q;R) &=&\gamma ^{\mu }+m_{f}^{2}\int \frac{d\Omega
_{s}}{4\pi }\frac{S^{\mu }S\hspace{-0.6229pc}/}{PS\,QS}\,;  \label{eff-ver-1}
\\
^{\ast }\widetilde{\Gamma }^{\mu \nu }(P,-P;K,-K) &=&-2m_{f}^{2}\int \frac{%
d\Omega _{s}}{4\pi }\frac{S^{\mu }S^{\nu }S\hspace{-0.6252pc}/}{PS\left(
P+K\right) \hspace{-2pt}S\left( P-K\right) \hspace{-2pt}S}\,.
\label{eff-ver-2}
\end{eqnarray}%
In both (\ref{eff-ver-1}) and (\ref{eff-ver-2}), $S\equiv (i,\mathbf{\hat{s}}%
)$ and $\Omega _{s}$ is the solid angle of $\mathbf{\hat{s}}$.

\subsection{Analytic expressions}

Let us recall the main results of \cite{ABD}. Henceforth, the quark thermal
mass $m_{f}$ is set to one. Since there remains another soft mass in the
problem, $m_{g}$, we define the ratio $m\left( N_{c},N_{f}\right) \equiv
m_{g}/m_{f}=\frac{4}{3}\sqrt{\frac{N_{c}(N_{c}+N_{f}/2)}{N_{c}^{2}-1}}$. It
is easy to see that we always have $m>1$. Since the gluon propagator is
taken in the strict Coulomb gauge, there are uncoupled longitudinal and
transverse contributions to $^{\ast }\Sigma _{1}$ and $^{\ast }\Sigma _{2}$.
There is a further split in $^{\ast }\Sigma _{1}$ due to positive and
negative helicity quarks. Six contributions in all, each being treated
separately.

After a preliminary manipulation of the gamma matrices, we perform the
expansion $\frac{1}{PS}=\frac{1}{ip_{0}}\left[ 1-\frac{\mathbf{p\hat{s}}}{%
ip_{0}}-\frac{\mathbf{p\hat{s}}^{2}}{p_{0}^{2}}+\mathcal{O}\left(
p^{3}\right) \right] $ in order to integrate analytically over the solid
angle $\Omega _{s}$. This expansion is valid in the region $p<\left\vert
ip_{0}\right\vert $, a condition always satisfied before analytic
continuation because $p_{0}=\left( 2n+1\right) \pi T$ and $p\sim g^{2}T$,
and after since for ultrasoft momenta, $ip_{0}=m_{f}+\mathcal{O}(\bar{p}%
)\sim gT$, see (\ref{omega-pm}). A subsequent expansion in powers of $p$ of
functions of $q=\left\vert \mathbf{p}-\mathbf{k}\right\vert $ is necessary
in order to carry out analytically the integration over the internal
three-momentum solid angle $\Omega _{k}$. In the process of these angular
integrations, special care must be taken when rotating the gamma matrices 
\cite{ABD}.

Next we perform the Matsubara sums. For this, the spectral decompositions of
the dressed propagators and $Q_{0}\left( K\right) $ with $k_{0}$ fermionic
are used \cite{pisarski-rho-1,pisarski-rho-2,le bellac}:%
\begin{eqnarray}
\Delta _{\varepsilon }(k_{0},k) &=&\int_{0}^{1/T}d\tau \,e^{ik_{0}\tau
}\int_{-\infty }^{+\infty }d\omega \,\rho _{\varepsilon }(\omega ,k)\left( 1-%
\tilde{n}(\omega )\right) e^{-\omega \tau }\,;  \notag \\
\Delta _{i}(k_{0},k) &=&\int_{0}^{1/T}d\tau \,e^{ik_{0}\tau }\int_{-\infty
}^{+\infty }d\omega \,\rho _{i}(\omega ,k)\left( 1+n(\omega )\right)
e^{-\omega \tau }\,;  \notag \\
Q_{0}(ik_{0}/k) &=&\int_{0}^{1/T}d\tau \,e^{ik_{0}\tau }\int_{-\infty
}^{+\infty }d\omega \,\rho _{0}(\omega ,k)\left( 1-\tilde{n}(\omega )\right)
e^{-\omega \tau }\,.  \label{spectral-relations}
\end{eqnarray}%
$\varepsilon $ stands for $+$ or $-$ and $i$ for $l$ or $t$. The functions $%
n(\omega )$ and $\tilde{n}(\omega )$ are the Bose-Einstein and Fermi-Dirac
distributions respectively, and the rho's are the spectral densities, to be
displayed in the next section. Before implementing the spectral
decomposition, it is first necessary to rearrange terms in such a way that
products of at most two functions necessitating a spectral decomposition
occur to ensure the obtainment of only one energy denominator just before
the extraction of the imaginary part. The steps of the calculation must also
ensure the Matsubara frequency $ik_{0}$ appears only in the numerator of
fractions. Hence, using (\ref{spectral-relations}), the sum over $k_{0}$ can
be performed, yielding a delta function that automatically removes one
integration over one imaginary time while the other integration produces an
energy denominator. At this stage, every $ip_{0}$ is to be replaced with $%
\left( 2n+1\right) \pi T$ except in the energy denominator.

Now the analytic continuation to real energies $ip_{0}\rightarrow \omega
_{\pm }(p)+i0^{+}$ can be taken. The extraction of the imaginary part
becomes straightforward using the relation $1/\left( x+i0^{+}\right) =\Pr
\left( 1/x\right) -i\pi \delta (x)$ where Pr stands for the principal part.
Further rearrangements are made and, according to the definitions in (\ref%
{gamma-pm-b}), the quark damping rates are given by the following
expressions:%
\begin{equation}
\gamma _{\pm }\left( p\right) =-\frac{g^{2}C_{f}T}{8\pi }\left[ a_{0}\pm
a_{1}\frac{\bar{p}}{3}+a_{2}\frac{\bar{p}^{2}}{9}+\mathcal{O}\left( \bar{p}%
^{3}\right) \right] .  \label{gam-pm-final}
\end{equation}%
Recall that we have set $m_{f}=1$. The coefficients $a_{i}$ are given by the
expressions:%
\begin{eqnarray}
a_{0} &=&\int_{\eta }^{\infty }dk\int_{-\infty }^{+\infty }d\omega
\int_{-\infty }^{+\infty }\frac{d\omega ^{\prime }}{\omega ^{\prime }}%
f_{0}\,\delta \,;  \notag \\
a_{1} &=&\int_{\eta }^{\infty }dk\int_{-\infty }^{+\infty }d\omega
\int_{-\infty }^{+\infty }\frac{d\omega ^{\prime }}{\omega ^{\prime }}\left[
\,f_{1}-\,f_{0}\,\partial _{\omega }\right] \delta ,  \notag \\
a_{2} &=&\int_{\eta }^{\infty }dk\int_{-\infty }^{+\infty }d\omega
\int_{-\infty }^{+\infty }\frac{d\omega ^{\prime }}{\omega ^{\prime }}%
\hspace{-2pt}\left[ f_{2}-f_{1}\partial _{\omega }-f_{0}\left( 3\,\partial
_{\omega }-\partial _{\omega }^{2}\right) \right] \delta ,  \label{a0-a1-a2}
\end{eqnarray}%
with $\delta =\delta \left( 1-\omega -\omega ^{\prime }\right) $. The three
functions $f_{i}\equiv f_{i}\left( \omega ,\omega ^{\prime };k\right) $ are
given by the following rather long expressions \cite{ABD}:%
\begin{equation}
f_{0}=\sum_{\varepsilon =\pm }\left[ -k^{2}\left( 1-\varepsilon k+\omega
\right) ^{2}\rho _{\varepsilon }\rho _{l}^{\prime }+\frac{1}{2}\left(
1+2\varepsilon k+k^{2}-\omega ^{2}\right) ^{2}\hspace{-2pt}\rho
_{\varepsilon }\rho _{t}^{\prime }\right] +\frac{1}{k}\left( k^{2}-\omega
^{2}\right) \rho _{0}\rho _{t}^{\prime }\,.  \label{f0}
\end{equation}%
This expression is the one obtained in \cite%
{kobes-kunstatter-mak,braaten-pisarski (quarks)} for the non-moving quark
damping rates. In the above expression and the two subsequent ones, the
notation is as follows: $\rho _{\varepsilon ,0}$ stands for $\rho
_{\varepsilon ,0}(\omega ,k)$ and $\rho _{l,t}^{\prime }$ for $\rho
_{l,t}(\omega ^{\prime },k)$. Remember also that $\partial _{\omega }$
stands for the partial derivative $\partial /\partial \omega $. The next
function $f_{1}\left( \omega ,\omega ^{\prime };k\right) $ is given by:%
\begin{eqnarray}
f_{1} &=&\sum_{\varepsilon =\pm }\left[ 2k^{2}\left( -1+k^{2}-2\varepsilon
k\omega +\omega ^{2}\right) \rho _{\varepsilon }\rho _{l}^{\prime }+\left[ -%
\frac{2\varepsilon }{k}-3+2\varepsilon k+4k^{2}-k^{4}\right. \right.  \notag
\\
&&\hspace{-28pt}\left. -\left( 2+4\varepsilon k+2k^{2}\right) \omega +\left( 
\frac{4\varepsilon }{k}+4+2\varepsilon k+2k^{2}\right) \omega ^{2}+2\omega
^{3}-\left( \frac{2\varepsilon }{k}+1\right) \omega ^{4}\right] \rho
_{\varepsilon }\rho _{t}^{\prime }  \notag \\
&&\hspace{-28pt}+\left. \hspace{-3pt}\varepsilon k^{2}\hspace{-2pt}\left( 1%
\hspace{-2pt}-\varepsilon k+\hspace{-2pt}\omega \right) ^{2}\hspace{-3pt}%
\rho _{\varepsilon }\partial _{k}\rho _{l}^{\prime }+\hspace{-1pt}\left[ 
\frac{\varepsilon }{2}+2k+3\varepsilon k^{2}+2k^{3}+\frac{\varepsilon }{2}%
k^{4}-\left( \varepsilon +2k+\varepsilon k^{2}\right) \allowbreak \omega
^{2}+\frac{\varepsilon }{2}\omega ^{4}\right] \allowbreak \hspace{-3pt}\rho
_{\varepsilon }\partial _{k}\rho _{t}^{\prime }\right]  \notag \\
&&\hspace{-28pt}-\frac{2}{k}\left( k^{2}-\omega ^{2}+2\frac{\omega ^{3}}{%
k^{2}}\right) \hspace{-3pt}\rho _{0}\rho _{t}^{\prime }-\hspace{-3pt}%
2k^{2}\epsilon \left( \omega \right) \delta \left( \omega ^{2}-k^{2}\right) 
\hspace{-3pt}\rho _{l}^{\prime }+\hspace{-3pt}\dfrac{\omega }{k^{2}}\hspace{%
-3pt}\left( \omega ^{2}-k^{2}\right) \hspace{-3pt}\rho _{0}\partial _{k}\rho
_{t}^{\prime }+2\omega \rho _{0}\partial _{k}\rho _{l}^{\prime },  \label{f1}
\end{eqnarray}%
where $\partial _{k}$ is a short notation for the partial derivative $%
\partial /\partial k$ and $\epsilon \left( \omega \right) $ is the sign
function. The last function $f_{2}\left( \omega ,\omega ^{\prime };k\right) $
to display is the longest of all three and is given by:%
\begin{eqnarray}
f_{2} &=&\sum_{\varepsilon =\pm }\hspace{-3pt}\left[ \left( -\frac{9}{2}%
-k^{2}-6\varepsilon k^{3}-\frac{1}{2}k^{4}-\left( 6\varepsilon
k-6k^{2}+2\varepsilon k^{3}\right) \allowbreak \omega +\left( 9+k^{2}\right)
\omega ^{2}+6\varepsilon k\omega ^{3}-\frac{9}{2}\omega ^{4}\right)
\allowbreak \right.  \notag \\
&&\hspace{-30pt}\times \rho _{\varepsilon }\rho _{l}^{\prime }+\hspace{-3pt}%
\left( \frac{9}{2k^{2}}-\frac{14\varepsilon }{k}-\hspace{-2pt}\frac{8}{3}+%
\hspace{-3pt}4\varepsilon k-\hspace{-2pt}\frac{19}{2}k^{2}\hspace{-2pt}-%
\hspace{-3pt}6\varepsilon k^{3}\hspace{-2pt}+\hspace{-2pt}k^{4}\allowbreak 
\hspace{-2pt}+\hspace{-3pt}\left( \frac{-3}{k^{2}}+\frac{25\varepsilon }{k}%
\hspace{-2pt}-\hspace{-2pt}10\hspace{-2pt}+\hspace{-3pt}6\varepsilon k%
\hspace{-2pt}+\hspace{-2pt}9k^{2}\hspace{-2pt}-\hspace{-3pt}3\varepsilon
k^{3}\hspace{-2pt}\right) \omega \right.  \notag \\
&&\hspace{-30pt}+\left( -\frac{6}{k^{2}}+\frac{2\varepsilon }{k}%
+23-6\varepsilon k+k^{2}\right) \allowbreak \omega ^{2}+\left( \frac{6}{k^{2}%
}-\frac{22\varepsilon }{k}-6+6\varepsilon k\right) \allowbreak \omega
^{3}+\left( -\frac{3}{2k^{2}}+\frac{12\varepsilon }{k}-5\right) \allowbreak
\omega ^{4}  \notag \\
&&\hspace{-30pt}-\left. \left( \frac{3}{k^{2}}+\frac{3\varepsilon }{k}%
\right) \allowbreak \omega ^{5}+\frac{3}{k^{2}}\allowbreak \omega
^{6}\right) \rho _{\varepsilon }\rho _{t}^{\prime }-k\left( 9-14\varepsilon
k+5k^{2}+\left( 12-2\varepsilon k-6k^{2}\right) \omega -\left(
3-12\varepsilon k\right) \omega ^{2}\right.  \notag \\
&&\hspace{-30pt}-\left. 6\omega ^{3}\right) \rho _{\varepsilon }\partial
_{k}\rho _{l}^{\prime }+\left( \frac{3}{2k}+4\varepsilon +7k+8\varepsilon
k^{2}+\frac{7}{2}k^{3}-\left( \frac{3}{k}+6\varepsilon +6k+6\varepsilon
k^{2}+3k^{3}\right) \omega \right.  \notag \\
&&\hspace{-30pt}-\hspace{-4pt}\left. \left( \frac{3}{k}+4\varepsilon
+5k\right) \hspace{-2pt}\omega ^{2}+\left( \frac{6}{k}+6\varepsilon
+6k\right) \hspace{-3pt}\omega ^{3}+\frac{3}{2k}\omega ^{4}-\frac{3}{k}%
\omega ^{5}\right) \rho _{\varepsilon }\partial _{k}\rho _{t}^{\prime }-%
\frac{3}{2}k^{2}\left( 1-\varepsilon k+\omega \right) ^{2}\rho _{\varepsilon
}\partial _{k}^{2}\rho _{l}^{\prime }  \notag \\
&&\hspace{-30pt}+\left. \left( \frac{3}{4}+3\varepsilon k+\frac{9}{2}%
k^{2}+3\varepsilon k^{3}+\frac{3}{4}k^{4}-\frac{3}{2}\left( 1+2\varepsilon
k+k^{2}\right) \omega ^{2}+\frac{3}{4}\omega ^{4}\right) \rho _{\varepsilon
}\partial _{k}^{2}\rho _{t}^{\prime }\right] \hspace{-3pt}-\frac{3}{k}\left(
k^{2}-\omega ^{2}\right) \rho _{0}\rho _{l}^{\prime }  \notag \\
&&\hspace{-30pt}+\left( \frac{3}{k}+2k+\frac{6}{k}\omega -\left( \frac{15}{%
k^{3}}+\frac{2}{k}\right) \omega ^{2}+\frac{18}{k^{3}}\omega ^{3}\right)
\rho _{0}\rho _{t}^{\prime }+\left( 6-12\omega \right) \rho _{0}\partial
_{k}\rho _{l}^{\prime }  \notag \\
&&\hspace{-30pt}+\left( -3+6k\omega +\frac{3}{k^{2}}\omega ^{2}-\frac{6}{k}%
\omega ^{3}\right) \rho _{0}\partial _{k}\rho _{t}^{\prime }+3k\rho
_{0}\partial _{k}^{2}\rho _{l}^{\prime }-\frac{3}{2k}\left( k^{2}-\omega
^{2}\right) \rho _{0}\partial _{k}^{2}\rho _{t}^{\prime }  \notag \\
&&\hspace{-30pt}+12k^{2}\epsilon \left( \omega \right) \delta \left( \omega
^{2}-k^{2}\right) \rho _{l}^{\prime }-6\omega \epsilon \left( \omega \right)
\delta \left( \omega ^{2}-k^{2}\right) \rho _{t}^{\prime }-6k^{2}\omega
\epsilon \left( \omega \right) \partial _{\omega ^{2}}\delta \left( \omega
^{2}-k^{2}\right) \rho _{l}^{\prime }\,.  \label{f2}
\end{eqnarray}%
Note that, since only soft values of $\omega $ and $\omega ^{\prime }$ are
to contribute, we have made use of the two approximations $\tilde{n}(\omega
)\simeq 1/2$ and $n(\omega )\simeq T/\omega $.

To complete the calculation of the damping rates, it remains to perform the
integrals over the frequencies $\omega $ and $\omega ^{\prime }$ and then
over the momentum $k.$ These integrations are not straightforward and
necessitate numerical work. But before that, one has to hunt down
divergences, particularly the infrared ones. Also, the dimensionless
parameter $m(N_{c},N_{f})$ is implicitly present in the spectral densities $%
\rho _{l,t}$ and so, each case $(N_{c},N_{f})$ has to be treated separately.
All this is discussed in the next section.

\section{Integration and divergences}

In this work, we will consider three cases regarding the number of colors
and flavors, namely, $(N_{c},N_{f})=\left( 2,1\right) $, $\left( 3,2\right) $
and $\left( 3,3\right) $. All the terms in the expressions of the
coefficients $a_{0}$, $a_{1}$ and $a_{2}$ in (\ref{gam-pm-final}) have the
following generic structure:%
\begin{equation}
I_{rs}=\int_{\eta }^{\infty }dk\int_{-\infty }^{+\infty }d\omega
\int_{-\infty }^{\infty }\frac{d\omega ^{\prime }}{\omega ^{\prime }}%
\,g\left( \omega ,k\right) \,\rho \left( \omega ,k\right) \,\partial
_{k}^{r}\rho ^{\prime }\left( \omega ^{\prime },k\right) \,\partial _{\omega
}^{s}\delta \left( 1-\omega -\omega ^{\prime }\right) ,  \label{Irs}
\end{equation}%
where $g\left( \omega ,k\right) $ is a polynomial in $\omega $ with
coefficients functions of $k$. The density $\rho \left( \omega ,k\right) $
stands for the quark spectral functions given in (\ref{rho-quark}) below or
for the spectral distributions given in (\ref{rho-0-1-2}). The density $\rho
^{\prime }\left( \omega ^{\prime },k\right) $ stands for the gluonic
spectral functions $\rho _{l,t}\left( \omega ^{\prime },k\right) $ given in (%
\ref{rho-gluon}). The two indices $r$ and $s$ designate the $r$th and $s$th
derivatives with respect to $k$ and $\omega $ respectively and are such that 
$r,s=0,1$ or 2 with the condition $r+s\leq 2$.

First we give the expressions of the spectral functions. The gluonic
spectral densities are as follows:%
\begin{equation}
\rho _{l,t}\left( \omega ,k\right) =\mathfrak{z}_{l,t}(k)\left[ \delta
\left( \omega -\omega _{l,t}(k)\right) -\delta \left( \omega +\omega
_{l,t}(k)\right) \right] +\beta _{l,t}\left( \omega ,k\right) \theta \left(
k-\left\vert \omega \right\vert \right) .  \label{rho-gluon}
\end{equation}%
The gluon energies $\omega _{l,t}(k)$ are the poles of the longitudinal and
transverse gluon propagators $^{\ast }\Delta _{l,t}\left( K\right) $ given
in (\ref{propa-glu}). The longitudinal and transverse residue functions are
given by:%
\begin{equation}
\mathfrak{z}_{l}(k)=\frac{\omega _{l}\left( k\right) \left( k^{2}-\omega
_{l}^{2}\left( k\right) \right) }{k^{2}\left( 3m^{2}-\omega _{l}^{2}\left(
k\right) +k^{2}\right) };\qquad \mathfrak{z}_{t}(k)=\frac{\omega _{t}\left(
k\right) \left( \omega _{t}^{2}\left( k\right) -k^{2}\right) }{3m^{2}\omega
_{t}^{2}\left( k\right) -\left( \omega _{t}^{2}\left( k\right) -k^{2}\right)
^{2}}.  \label{z-gluon}
\end{equation}%
The longitudinal cut function is given by:%
\begin{equation}
\beta _{l}\left( k,\omega \right) =\frac{-3m^{2}\omega }{2k\left[ \left(
3m^{2}+k^{2}-3m^{2}\frac{\omega }{2k}\ln \left( \frac{k+\omega }{k-\omega }%
\right) \right) ^{2}+\left( \frac{3\pi m^{2}\omega }{2k}\right) ^{2}\right] }%
,  \label{beta-long}
\end{equation}%
and the transverse one by:%
\begin{equation}
\beta _{t}\left( k,\omega \right) =\frac{3m^{2}\omega (k^{2}-\omega ^{2})}{%
2k^{3}\left[ \left( \omega ^{2}-k^{2}-\frac{3}{2}m^{2}\left( \frac{\omega
^{2}}{k^{2}}+\frac{\omega \left( k^{2}-\omega ^{2}\right) }{2k^{3}}\ln
\left( \frac{k+\omega }{k-\omega }\right) \right) \right) ^{2}+\left( \frac{%
3\pi m^{2}\omega \left( k^{2}-\omega ^{2}\right) }{4k^{3}}\right) ^{2}\right]
}.  \label{beta-trans}
\end{equation}%
The quark spectral functions are:%
\begin{equation}
\rho _{\pm }\left( \omega ,k\right) =\mathfrak{z}_{\pm }(k)\delta \left(
\omega -\omega _{\pm }(k)\right) +\mathfrak{z}_{\mp }(k)\delta \left( \omega
+\omega _{\pm }(k)\right) +\beta _{\pm }\left( \omega ,k\right) \theta
\left( k-\left\vert \omega \right\vert \right) ,  \label{rho-quark}
\end{equation}%
where the quark residue functions are:%
\begin{equation}
\mathfrak{z}_{\pm }(k)=-\frac{1}{2}\left( \omega _{\pm }^{2}\left( k\right)
-k^{2}\right) ,  \label{z-quark}
\end{equation}%
and the quark cut functions given by:%
\begin{equation}
\beta _{\pm }\left( \omega ,k\right) =\dfrac{-(k\mp \omega )}{2k^{2}\left[
\left( \omega \mp k+\frac{1}{2k^{2}}\left( \left( k\mp \omega \right) \ln
\left( \frac{k\pm \omega }{k\mp \omega }\right) +2k\right) \right)
^{2}+\left( \frac{\pi \left( k\mp \omega \right) }{2k^{2}}\right) ^{2}\right]
}.  \label{beta-quark}
\end{equation}%
The other spectral distributions appearing in the functions $f_{0}$, $f_{1}$
and $f_{2}$ are as follows:%
\begin{equation}
\rho _{0}\left( \omega ,k\right) \hspace{-2pt}=\hspace{-2pt}\tfrac{-1}{2}%
\theta \left( k^{2}-\omega ^{2}\right) ;\quad \rho _{1}\left( \omega
,k\right) \hspace{-2pt}=\hspace{-2pt}\epsilon \left( \omega \right) \delta
\left( k^{2}-\omega ^{2}\right) ;\quad \rho _{2}\left( \omega ,k\right) 
\hspace{-2pt}=\hspace{-2pt}\epsilon \left( \omega \right) \partial _{\omega
^{2}}\delta \left( k^{2}-\omega ^{2}\right) .  \label{rho-0-1-2}
\end{equation}

To show how to carry with the generic integral $I_{rs}$, it is appropriate
to chose the spectral function $\rho \left( \omega ,k\right) $ as a quark
spectral density. This example is typical and general enough to encompasses
all the major difficulties and subtleties we encounter throughout the work.
Using the expressions in (\ref{rho-quark}) and (\ref{rho-gluon}), we see
that there are four kinds of contributions: pole-pole ($pp$), pole-cut ($pc$%
), cut-pole ($cp$) and cut-cut ($cc$). Let us start with the pole-pole
contribution, which typically writes as:%
\begin{equation}
I_{rs}^{pp}=\int_{\eta }^{\infty }\hspace{-4pt}dk\int_{-\infty }^{\infty }%
\hspace{-4pt}d\omega \int_{-\infty }^{\infty }\hspace{-4pt}\frac{d\omega
^{\prime }}{\omega ^{\prime }}g\left( \omega ,k\right) \mathfrak{z}_{\pm
\varepsilon }\left( k\right) \delta \left( \omega \mp \omega _{\varepsilon
}\right) \partial _{k}^{r}\left[ \mathfrak{z}_{l,t}\left( k\right) \delta
\left( \omega ^{\prime }\pm \omega _{l,t}\right) \right] \hspace{-2pt}%
\partial _{\omega }^{s}\delta \left( 1-\omega -\omega ^{\prime }\right) .
\label{Irs-pp}
\end{equation}%
The integrand is nonzero only at the intersections of the supports of the
delta functions, namely, when $\omega \pm \omega _{\varepsilon }=0$, $\omega
^{\prime }\pm \omega _{l,t}=0$ and $\omega +\omega ^{\prime }=1$. This is
satisfied at the points $k$ for which we have $\pm \omega _{\varepsilon
}\left( k\right) =1\mp \omega _{l,t}\left( k\right) $. But for the three
cases $\left( N_{c},N_{f}\right) $ considered in this work, these curves
never intersect. Hence we can write:%
\begin{equation}
I_{rs}^{pp}=0.  \label{Irs-pp final}
\end{equation}

Next we look at the term pole-cut. It is of the form:%
\begin{equation}
I_{rs}^{pc}=\int_{\eta }^{\infty }\hspace{-6pt}dk\int_{-\infty }^{\infty }%
\hspace{-6pt}d\omega \int_{-\infty }^{\infty }\hspace{-4pt}\frac{d\omega
^{\prime }}{\omega ^{\prime }}g\left( \omega ,k\right) \mathfrak{z}%
_{\varepsilon }\delta \left( \omega \pm \omega _{\varepsilon }\right)
\partial _{k}^{\left( r\right) }\hspace{-2pt}\left[ \beta _{l,t}\left(
\omega ^{\prime },k\right) \theta \left( k-\left\vert \omega ^{\prime
}\right\vert \right) \right] \partial _{\omega }^{\left( s\right) }\delta
\left( 1-\omega -\omega ^{\prime }\right) .  \label{Irs-pc}
\end{equation}%
With the aim of making the illustration clear, let us consider first the
specific case $r=1$ and $s=0$. The integration over $\omega ^{\prime }$
becomes trivial. The derivation with respect to $k$ yields two terms: $%
g\left( \omega ,k\right) \mathfrak{z}_{\varepsilon }\left( k\right) \partial
_{k}\beta _{l,t}\left( 1-\omega ,k\right) \delta \left( \omega \pm \omega
_{\varepsilon }\right) \theta \left( k-\left\vert 1-\omega \right\vert
\right) $ and $g\left( \omega ,k\right) \mathfrak{z}_{\varepsilon }\left(
k\right) \beta _{l,t}\left( 1-\omega ,k\right) \delta \left( \omega \pm
\omega _{\varepsilon }\right) \delta \left( k-\left\vert 1-\omega
\right\vert \right) $. The second term is always zero because it imposes the
conditions $\omega _{\varepsilon }\left( k\right) =\pm \left( 1\mp k\right) $%
, which are never satisfied kinematically. In the first term, the
integration over $\omega $ is straightforward and the theta function imposes
the constraints $-k\leq 1\pm \omega _{\varepsilon }\left( k\right) \leq k$
on the momentum integration. Only the minus sign can be satisfied, and for
all values of $k$. Hence we write for this contribution:%
\begin{equation}
I_{10}^{pc}=\int_{\eta }^{\infty }dk\frac{g\left( \omega _{\varepsilon
},k\right) }{1-\omega _{\varepsilon }\left( k\right) }\mathfrak{z}%
_{\varepsilon }\left( k\right) \left. \partial _{k}\beta _{l,t}\left( \omega
,k\right) \right\vert _{\omega =1-\omega _{\varepsilon }\left( k\right) }.
\label{I10-pc}
\end{equation}%
The case $r=2$ and $s=0$ is carried out in a similar way. Because of the
second derivative in $k$, there will be three contributions: (i) $g\left(
\omega ,k\right) \mathfrak{z}_{\varepsilon }\left( k\right) \partial
_{k}^{2}\beta _{l,t}\left( 1-\omega ,k\right) \delta \left( \omega \pm
\omega _{\varepsilon }\right) \theta \left( k-\left\vert 1-\omega
\right\vert \right) $. The integration over $\omega $ is trivial and the
theta function imposes the minus sign with no constraints on $k$. (ii) 2$%
g\left( \omega ,k\right) \mathfrak{z}_{\varepsilon }\left( k\right) \partial
_{k}\beta _{l,t}\left( 1-\omega ,k\right) \delta \left( \omega \pm \omega
_{\varepsilon }\right) \delta \left( k-\left\vert 1-\omega \right\vert
\right) $ and (iii) $g\left( \omega ,k\right) \mathfrak{z}_{\varepsilon
}\left( k\right) \beta _{l,t}\left( 1-\omega ,k\right) \delta \left( \omega
\pm \omega _{\varepsilon }\right) \partial _{k}\delta \left( k-\left\vert
1-\omega \right\vert \right) $. These two contributions cancel because of
the same kinematics as in the previous case. Then we can write:%
\begin{equation}
I_{20}^{pc}=\int_{\eta }^{\infty }dk\frac{g\left( \omega _{\varepsilon
},k\right) }{1-\omega _{\varepsilon }\left( k\right) }\mathfrak{z}%
_{\varepsilon }\left( k\right) \left. \partial _{k}^{2}\beta _{l,t}\left(
\omega ,k\right) \right\vert _{\omega =1-\omega _{\varepsilon }\left(
k\right) }.  \label{I20-pc}
\end{equation}%
The other cases for $r$ and $s$ are worked out in similar steps. When a
derivative with respect to $\omega $ intervenes over $\delta \left( 1-\omega
-\omega ^{\prime }\right) $, it is first transformed into an integral over $%
\omega ^{\prime }$ and then brought back onto the other functions. The rest
of the steps are straightforward. We can write generically:%
\begin{equation}
I_{rs}^{pc}=\left( -\right) ^{s}\int_{\eta }^{\infty }dk\,g\left( \omega
_{\varepsilon },k\right) \mathfrak{z}_{\varepsilon }\left( k\right) \left.
\partial _{k,\omega }^{r,s}\frac{\beta _{l,t}\left( \omega ,k\right) }{%
\omega }\right\vert _{\omega =1-\omega _{\varepsilon }\left( k\right) }.
\label{Irs-pc final}
\end{equation}%
The notation $\partial _{k,\omega }^{r,s}$ means derive $r$-times with
respect to $k$ and $s$-times with respect to $\omega $.

\subsection{Infrared behavior}

The generic integral $I_{rs}^{pc}$ is present in all three coefficients $%
a_{0}$, $a_{1}$ and $a_{2}$. From the explicit expression of $\mathfrak{z}%
_{\varepsilon }\left( k\right) $ given in (\ref{z-quark}) and those of $%
\beta _{l,t}\left( \omega ,k\right) $ given in (\ref{beta-long}) and (\ref%
{beta-trans}), one carries out the $k$-integration in (\ref{Irs-pc final})
for the different specific functions $g\left( \omega ,k\right) $ intervening
in the different contributions to $f_{0}$, $f_{1}$ and $f_{2}$ given in
relations (\ref{f0}), (\ref{f1}) and (\ref{f2}) respectively. Most of the
work is numerical. For the coefficients $a_{0}$ and $a_{1}$, no particular
difficulty arises and the limit $\eta \rightarrow 0$ is safe. However, the
coefficient $a_{2}$ requires special attention since the infrared limit is
sensitive. To see this concretely, take the explicit example of $\rho
_{+}\partial _{k}^{2}\rho _{l}^{\prime }$ in $f_{2}$ with $g\left( \omega
,k\right) =-3k^{2}(1-k+\omega )^{2}$. Let us write the integral as $%
I_{20}^{pc}=\int_{\eta }^{\infty }dkF\left( k\right) $ with the integrand $%
F\left( k\right) =-3k^{2}(1-k+\omega _{+})^{2}\mathfrak{z}_{+}/\left(
1-\omega _{+}\right) \left. \partial _{k}^{2}\beta _{l}\left( \omega
,k\right) \right\vert _{\omega =1-\omega _{+}}$. With the explicit
expressions of $\omega _{+}\left( k\right) $ given in (\ref{omega-pm}), $%
\mathfrak{z}_{+}\left( k\right) $ in (\ref{z-quark}) and $\beta _{l}\left(
\omega ,k\right) $ in (\ref{beta-long}), we can perform a small-$k$
expansion of the integrand $F\left( k\right) $ to find for the different
cases of $N_{c}$ and $N_{f}$:%
\begin{align}
\left( N_{c};N_{f}\right) & =\left( 2,1\right) \longrightarrow F\left(
k\right) =-\dfrac{0.4486}{k}+0.4486+0.2248k+\mathcal{O}\left( k^{2}\right) ;
\notag \\
\left( N_{c};N_{f}\right) & =\left( 3,2\right) \longrightarrow F\left(
k\right) =-\dfrac{0.4985}{k}+0.5428+0.2466k+\mathcal{O}\left( k^{2}\right) ;
\notag \\
\left( N_{c};N_{f}\right) & =\left( 3,3\right) \longrightarrow F\left(
k\right) =-\dfrac{0.4431}{k}+0.4825+0.2223k+\mathcal{O}\left( k^{2}\right) .
\label{expansion F}
\end{align}%
The $1/k$ behavior indicates a logarithmic infrared divergence. This latter
is extracted in the following manner. Writing $F\left( k\right) =\alpha /k+%
\mathrm{finite}$, we have:%
\begin{equation}
I_{20}^{pc}=-\alpha \ln \eta +\alpha \ln \lambda +\int_{0}^{\lambda
}dk\left( F\left( k\right) -\dfrac{\alpha }{k}\right) +\int_{\lambda
}^{\infty }dkF\left( k\right) .  \label{I20-pc example}
\end{equation}%
In short, we have split the original integral into two: one from $\eta $ to
an arbitrary number $\lambda $ and another one from $\lambda $ to $\infty $.
The second integral is finite. From the integrand of the first one we have
subtracted $\alpha /k$, which makes the integral safe in the infrared and
the limit $\eta \rightarrow 0$ can be taken. The term $\alpha /k$ has to be
integrated by itself from $\eta $ to $\lambda $. As said, the choice of $%
\lambda $ is arbitrary and we must (and do) check that the result in
independent of it. For this specific example, we obtain the following
results:%
\begin{align}
\left( N_{c};N_{f}\right) & =\left( 2,1\right) \longrightarrow
I_{20}^{pc}=0.4486\ln \eta -0.7641;  \notag \\
\left( N_{c};N_{f}\right) & =\left( 3,2\right) \longrightarrow
I_{20}^{pc}=0.4985\ln \eta -0.8557;  \notag \\
\left( N_{c};N_{f}\right) & =\left( 3,3\right) \longrightarrow
I_{20}^{pc}=0.4431\ln \eta -0.7539.  \label{I20-pc example final}
\end{align}%
All other infrared-sensitive contributions are worked out in a similar
manner.

\subsection{Other circumvented singularities}

The upcoming cut-pole contribution is not sensitive to $\eta $ but has other
difficulties we discuss now. Consider then:%
\begin{equation}
I_{rs}^{cp}=\int_{\eta }^{\infty }dk\int_{-\infty }^{\infty }d\omega
\int_{-\infty }^{\infty }\frac{d\omega ^{\prime }}{\omega ^{\prime }}g\left(
\omega ,k\right) \beta _{\varepsilon }\left( \omega ,k\right) \theta \left(
k-\left\vert \omega \right\vert \right) \partial _{k}^{r}\left[ \mathfrak{z}%
_{i}\delta \left( \omega ^{\prime }-\omega _{i}\right) \right] \partial
_{\omega }^{s}\delta \left( 1-\omega -\omega ^{\prime }\right) ,
\label{Irs-cp}
\end{equation}%
where $i$ stands for $l$ or $t$. Let us be a little more specific and
consider the case $r=2$ and $s=0$. The integration over $\omega ^{\prime }$
is trivial. We obtain:%
\begin{equation}
I_{20}^{cp}=\int_{\eta }^{\infty }dk\int_{-\infty }^{\infty }\frac{d\omega }{%
1-\omega }\left[ g\left( \omega ,k\right) \beta _{\varepsilon }\left( \omega
,k\right) \theta \left( k-\left\vert \omega \right\vert \right) \left[ 
\mathfrak{z}_{i}^{\prime \prime }+2\mathfrak{z}_{i}^{\prime }\partial _{k}+%
\mathfrak{z}_{i}\partial _{k}^{2}\right] \delta \left( 1-\omega -\omega
_{i}\right) \right] .  \label{I20-cp}
\end{equation}%
Here, $\mathfrak{z}_{i}^{\prime }\left( k\right) $ is the first derivative
and $\mathfrak{z}_{i}^{\prime \prime }\left( k\right) $ the second
derivative of the residue function. The kinematics imposes $1-k\leq \omega
_{i}\left( k\right) \leq 1+k$, conditions satisfied for all $k$ from the
lower bound $k_{i}$ to $\infty ,$ where $k_{i}$ is the solution to the
condition $\omega _{i}\left( k\right) =1+k$. In the last two contributions,
the $k$-derivatives over the delta function are brought onto the other
functions. With some algebra and using the fact that $\lim_{\omega
\rightarrow \pm k}\beta _{\varepsilon }\left( \omega ,k\right) =0$ to
eliminate some of the intermediary terms, we can write:%
\begin{eqnarray}
I_{20}^{cp} &=&\int_{k_{i}}^{\infty }dk\,\left[ \mathfrak{z}_{i}^{\prime
\prime }-\left( \mathfrak{z}_{i}\omega _{i}^{\prime \prime }+2\mathfrak{z}%
_{i}^{\prime }\omega _{i}^{\prime }\right) \partial _{\omega }+\mathfrak{z}%
_{i}\omega _{i}^{\prime }{}^{2}\partial _{\omega }^{2}\right] \left[ g\left(
\omega ,k\right) \beta _{\epsilon }\left( \omega ,k\right) /\left( 1-\omega
\right) \right] _{\omega =1-\omega _{i}}  \notag \\
&&+\left[ \frac{\mathfrak{z}_{i}\omega _{i}^{\prime }{}^{2}g\left( 1-\omega
_{i},k\right) }{\omega _{i}\mid 1-\omega _{i}^{\prime }\mid }\left[ 2\left.
\partial _{\omega }\beta _{\epsilon }\left( \omega ,k\right) \right\vert
_{\omega =1-\omega _{i}}-\dfrac{\beta _{\epsilon }^{\prime }\left( 1-\omega
_{i},k\right) }{\mid 1-\omega _{i}^{\prime }\mid }\right] \right] _{k=k_{i}}.
\label{I20-cp 2nd expression}
\end{eqnarray}%
In this expression, $\beta _{\epsilon }^{\prime }\left( 1-\omega
_{i},k\right) $ indicates the total derivative with respect to $k$. Now we
must be careful with this expression because the derivatives of the quark
cut functions at $k_{i}$ are infinite. This is not related to the infrared
limit. However, putting these infinities together cancels them. More
specifically, if we call $y=k-k_{i}$, then the singular behavior of the
integrand around $k_{i}$ coming from the derivatives of the quark cut
functions has the form $A\left( \ln y,y\right) /y^{2}+B\left( \ln y,y\right)
/y$, where $A$ and $B$ are complicated but computable functions. The task is
then to systematically extract this singular behavior and put it together
with a similar one coming from the terms not under the integral sign. It
turns out that in each case and for every term, when put together, all the
singularities cancel.

\subsection{Light-cone behavior}

Last we turn to the cut-cut contribution to $I_{rs}$. Here a different kind
of potential singularities arises, light-cone, but circumvented too. The
generic term has the following form:%
\begin{eqnarray}
I_{rs}^{cc} &=&\int_{\eta }^{\infty }dk\int_{-\infty }^{\infty }d\omega
\int_{-\infty }^{\infty }\frac{d\omega ^{\prime }}{\omega ^{\prime }}g\left(
\omega ,k\right) \beta _{\varepsilon }\left( \omega ,k\right) \theta \left(
k-\left\vert \omega \right\vert \right)   \notag \\
&&\times \partial _{k}^{r}\left[ \beta _{i}(\omega ^{\prime },k)\theta
\left( k-\left\vert \omega ^{\prime }\right\vert \right) \right] \partial
_{\omega }^{m}\delta \left( 1-\omega -\omega ^{\prime }\right) .
\label{Irs-cc}
\end{eqnarray}%
Here too the case $r=2$ and $s=0$ is typical enough. Performing the integral
over $\omega ^{\prime }$ trivially, we have:%
\begin{align}
I_{20}^{cc}& =\int_{0}^{\infty }dk\int_{-\infty }^{\infty }\frac{d\omega }{%
1-\omega }g\left( \omega ,k\right) \beta _{\varepsilon }\left( \omega
,k\right) \theta \left( k-\left\vert \omega \right\vert \right)   \notag \\
& \times \left[ \partial _{k}^{2}\beta _{i}(1-\omega ,k)\,\theta \left(
k-\left\vert 1-\omega \right\vert \right) +2\partial _{k}\beta _{i}(1-\omega
,k)\,\delta \left( k-\left\vert 1-\omega \right\vert \right) \right.   \notag
\\
& +\left. \beta _{i}(1-\omega ,k)\partial _{k}\delta \left( k-\left\vert
1-\omega \right\vert \right) \right] .  \label{I20-cc}
\end{align}%
The $\Theta \Theta $ term is constrained by $1-k\leq \omega \leq k$ with $%
k\geq 0.5$. We obtain the contribution $\int_{0.5}^{\infty
}dk\int_{1-k}^{k}d\omega g\left( \omega ,k\right) \beta _{\varepsilon
}\left( \omega ,k\right) \partial _{k}^{2}\beta _{i}(1-\omega ,k)/\left(
1-\omega \right) $. The $\Theta \delta $ contribution is left with one
integral over $k$; it reads $2\int_{0.5}^{\infty }\frac{dk}{k}g\left(
1-k,k\right) \beta _{\varepsilon }\left( 1-k,k\right) \left. \partial
_{k}\beta _{i}(1-\omega ,k)\right\vert _{\omega =1-k}$. The $\Theta \partial
_{k}\delta $ contribution is treated similarly and yields $%
-\int_{0.5}^{\infty }\frac{dk}{k}\partial _{k}\left[ g\left( \omega
,k\right) \beta _{\varepsilon }\left( \omega ,k\right) \beta _{i}(1-\omega
,k)\right] _{\omega =1-k}$. However, it is easy to see that we can get rid
of the second and third contributions by performing one integration by part
in the first contribution. It will yield $-\int_{0.5}^{\infty
}dk\int_{1-k}^{k}\frac{d\omega }{1-\omega }\partial _{k}\left[ g\left(
\omega ,k\right) \beta _{\varepsilon }\left( \omega ,k\right) \right]
\partial _{k}\beta _{i}\left( 1-\omega ,k\right) $ plus the opposite of the
second and third contributions. Hence, changing the integration from over $%
\omega $ to over $1-\omega $, we can write:

\begin{equation}
I_{20}^{cc}=-\int_{0.5}^{\infty }dk\int_{1-k}^{k}\frac{d\omega }{\omega }%
\partial _{k}\left[ g\left( 1-\omega ,k\right) \beta _{\varepsilon }\left(
1-\omega ,k\right) \right] \partial _{k}\beta _{i}\left( \omega ,k\right) .
\label{I20-cc 2nd expression}
\end{equation}

But extra care must be taken because $\lim_{\omega \rightarrow k}\partial
_{k}\beta _{i}\left( \omega ,k\right) $ in infinite. This situation is
handled as follows. Given the explicit expressions (\ref{beta-long}) and (%
\ref{beta-trans}) of the longitudinal and transverse cut functions
respectively, one can see that the divergence in $\partial _{k}\beta
_{i}\left( \omega ,k\right) $ comes from $1/\zeta \equiv 1/\left( k-\omega
\right) $ and $Y^{-1}\equiv \ln \zeta $. Then, writing $I_{20}^{cc}=%
\int_{0.5}^{\infty }dk\int_{0}^{2k-1}d\zeta G\left( k,\zeta ;Y\right) $, we
expand the integrand $G\left( k,\zeta ;Y\right) $ in powers of $\zeta $: 
\begin{equation}
G\left( k,\zeta ;Y\right) =G_{-1}\left( k,Y\right) /\zeta +G_{0}\left(
k,Y\right) +G_{1}\left( k,Y\right) \zeta +\mathcal{O}\left( \zeta
^{2}\right) .  \label{expansion of G}
\end{equation}%
The singular term $G_{-1}\left( k,Y\right) /\zeta $ is singled out in the
following manner:%
\begin{equation}
I_{20}^{cc}=\int_{0.5}^{+\infty }\hspace{-4pt}dk\int_{0}^{2k-1}\hspace{-4pt}%
d\zeta \left[ G\left( k,\zeta ;1/\ln \zeta \right) -G_{-1}\left( k,1/\ln
\zeta \right) /\zeta \right] +\hspace{-4pt}\int_{0.5}^{\infty }\hspace{-4pt}%
dk\int_{\ln ^{-1}\left( 2k-1\right) }^{0^{-}}\hspace{-4pt}\dfrac{dY}{Y^{2}}%
G_{-1}(k,Y).  \label{I20-cc 3rd expression}
\end{equation}%
In this expression, the first integral is finite. In the second integral,
knowing that $G_{-1}(k,Y)=\mathcal{O}\left( Y^{3}\right) $ close to zero,
there is no divergence anymore. Hence the whole $I_{20}^{cc}$ is finite. The
same trick is used for all the cut-cut terms sensitive close to the
light-cone and all singularities are circumvented.

\section{Results and conclusion}

All the terms in $f_{0}$, $f_{1}$ and $f_{2}$ given in (\ref{f0}), (\ref{f1}%
) and (\ref{f2}) respectively have to be calculated. There are no additional
subtleties to mention. Assembling all the partial results together and
reintroducing the quark thermal mass $m_{f}$ such that $\bar{p}=p/m_{f}$ and 
$\bar{\eta}=\eta /m_{f}$, we find the following final results:%
\begin{align}
\left( N_{c};N_{f}\right) & =\left( 2,1\right) \longrightarrow \gamma _{\pm
}\left( p\right) =\dfrac{3g^{2}T}{64\pi }\hspace{-2pt}\left[ 5.6978\mp 1.0452%
\bar{p}-\hspace{-2pt}(6.0427\ln \bar{\eta}-8.4684)\bar{p}^{2}+\hspace{-2pt}%
\mathcal{O}\left( \bar{p}^{3}\right) \right] ;  \notag \\
\left( N_{c};N_{f}\right) & =\left( 3,2\right) \longrightarrow \gamma _{\pm
}\left( p\right) =\dfrac{g^{2}T}{12\pi }\left[ 5.6344\mp 0.9492\bar{p}%
-(6.7141\ln \bar{\eta}-7.7539)\bar{p}^{2}+\mathcal{O}\left( \bar{p}%
^{3}\right) \right] ;  \notag \\
\left( N_{c};N_{f}\right) & =\left( 3,3\right) \longrightarrow \gamma _{\pm
}\left( p\right) =\dfrac{g^{2}T}{12\pi }\left[ 5.7057\mp 1.0568\bar{p}%
-(5.9681\ln \bar{\eta}-8.5536)\bar{p}^{2}+\mathcal{O}\left( \bar{p}%
^{3}\right) \right] .  \label{gamma-pm}
\end{align}

Remember that the present calculation of the quark damping rates is done in
the context of next-to-leading order hard-thermal-loop summed perturbative
QCD at high temperature where weak coupling is assumed. It is intended to be
an additional contribution to probe the analytic properties of
finite-temperature QCD, particularly in the infrared. As mentioned in the
introductory remarks, it has no direct relevance to the physics of the
strongly coupled quark-gluon plasma, but it may be useful if, eventually,
quantum chromodynamics is still believed to be the bedrock of all
theoretical hadronic physics.

One peculiarity of the results (\ref{gamma-pm}) is that the logarithmic
sensitivity to the infrared is found only in the coefficients of $p^{2}$;
the zeroth and first order coefficients are safe. The calculations have
encountered other potential divergences: at the light-cone and at specific
points determined by the gluon energies. But all these have been
circumvented.

The persistence of the infrared divergence is most probably attributed to
the non-screening of the static chromo-magnetic fields at lowest order. In a
separate work \cite{AB}, we have considered the context of scalar quantum
electrodynamics, where calculations are more fluid, and have explicitly
shown that an early momentum expansion like the one we perform in \cite{ABD}
and this work gives exactly the same results as a late expansion, after the
Matsubara sum and analytic continuation to real energies are done. Remember
that only soft values of the internal momentum have been included in the
integration. It seems then that the ultrasoft region is important and needs
to be investigated with probably non-perturbative means.

\end{document}